\documentclass[preprint,12pt]{elsarticle}

\usepackage{amssymb}
\usepackage{amsmath}

\usepackage{lineno}

\usepackage{booktabs}

\usepackage{hyperref}

\usepackage[left=1.05in,right=1.05in]{geometry}

\journal{Journal of Computational Physics}

\begin{document}

\begin{frontmatter}



\title{Tensor Network Compression for Fully Spectral Vlasov--Poisson Simulation}


\author{Erik M. Åsgrim, Luca Pennati, Marco Pasquale, Stefano Markidis} 

\affiliation{organization={Department of Computational Science and Technology, KTH Royal Institute of Technology},
            addressline={Lindstedtsvägen 5},
            city={Stockholm},
            postcode={114 28},
            country={Sweden}}

\begin{abstract}
We propose a numerical method for kinetic plasma simulation in which the phase-space distribution is represented by a low-rank tensor network with an adaptive level of compression. The Vlasov--Poisson system is advanced using a second-order Strang splitting scheme, with the advection and acceleration steps treated spectrally in position and velocity, respectively. By representing both the state and the operators required for time evolution as compressed tensor objects, the propagation can be carried out directly in tensor form without reconstructing the full phase-space grid. The self-consistent electric field is likewise obtained entirely within the tensor formalism through a tensor-based Poisson solver. We validate the approach on Landau damping and the two-stream instability, and find near-indistinguishable agreement with the corresponding full-grid reference while achieving substantial compression of the state representation. Additionally, we examine how varying the degree of compression influences conservation properties, positivity behavior, and computational cost, and show that compressibility varies strongly across the considered dynamical regimes. More broadly, the results point toward a promising class of spectral Eulerian plasma solvers that operate directly on compressed tensor representations.

\end{abstract}




\end{frontmatter}

\newpage
\section{Introduction}
Plasma simulation models are essential for understanding and predicting plasma dynamics across a wide range of settings, from space and astrophysical environments~\cite{kulsrud2005plasma_astrophysics} to magnetic confinement fusion~\cite{hazeltine2003plasma_confinement} and accelerator-based technologies~\cite{chen2016intro_plasma}. Depending on the physical regime and desired level of accuracy, a variety of numerical descriptions are used in practice. Fluid models treat the plasma as a continuum and evolve macroscopic moments such as density, momentum, and energy~\cite{goedbloed2004principles_mhd,goossens_introduction_2003}, whereas kinetic models provide a more fundamental description by evolving the phase-space distribution function, which represents the statistical density of particles as a function of position and velocity~\cite{nicholson1983introduction}. In many regimes of interest, collisional effects are negligible on the considered timescales, such that the distribution function evolves according to the collisionless Vlasov equation. The resulting plasma dynamics are coupled self-consistently to electromagnetic fields governed by Maxwell's equations, or by the Poisson equation in the electrostatic limit~\cite{chen2016intro_plasma}. This coupling is inherently nonlinear, as the force term in the Vlasov equation depends on fields determined by moments of the distribution function, while those moments are themselves obtained by integrating the evolving distribution over velocity.

A primary numerical challenge in kinetic plasma simulation is the cost of representing the phase-space distribution function~\cite{heath_discontinuous_2012,einkemmer_low-rank_2020}. In a standard Eulerian discretization, both physical space and velocity space are sampled on grids~\cite{sonnendrucker_semi-lagrangian_1999,filbet_conservative_2001}. While spatial resolution can often be managed with established discretization and parallelization strategies, accurately resolving velocity-space structure is substantially more demanding. Phase-space mixing and filamentation generate increasingly fine features that require very high resolution~\cite{sonnendrucker_semi-lagrangian_1999,filbet_conservative_2001}. The situation becomes prohibitive in the full six-dimensional phase space, where the number of degrees of freedom grows exponentially with dimension, exemplifying the \emph{curse of dimensionality}~\cite{einkemmer_mass_2021}. For this reason, many Vlasov solvers employ adaptive or problem-dependent strategies for velocity-space discretization~\cite{gutnic_vlasov_2004,hittinger_block-structured_2013}. An alternative is to introduce a compressed representation of the distribution function and formulate the governing equations directly in that compressed space, avoiding repeated encoding and decoding of a full grid-based state~\cite{einkemmer_low-rank_2018}. Particle-in-Cell (PIC) methods can be viewed as one such compression strategy, in which the distribution is represented by a finite set of macroparticles, and the kinetic dynamics are advanced along characteristics~\cite{birdsall_plasma_1985,hockney_computer_1988}. This representation is inherently lossy and introduces sampling noise, so that regions of phase space that are poorly populated by particles may be inaccurately represented, and small-amplitude features can be missed without very large particle counts.

In this study, we pursue an alternative compression paradigm based on low-rank factorization of the phase-space distribution function~\cite{koch_dynamical_2007,einkemmer_low-rank_2018,oseledets_tensor-train_2011}. Specifically, we adopt techniques originally developed for the numerical simulation of many-body quantum systems and represent the distribution function using a tensor network. This approach offers three key advantages. First, tensor networks provide a compressed low-rank representation of large-scale phase-space data, with the potential to reduce storage and computational costs by orders of magnitude when the underlying solution admits a suitable low-rank structure. Although no general guarantees exist that a given Vlasov solution admits such low-rank structure uniformly in time, a growing body of empirical evidence indicates that many physically relevant kinetic problems possess an intrinsic, approximately low-rank structure in appropriate representations~\cite{kormann_semi-lagrangian_2015, gourianov_quantum_2022, ishida_low-rank_2025, dolgov_low-rank_2014, niedermeier_solving_2025}. Second, the tensor network framework allows the degree of compression to be tuned adaptively, ranging from essentially lossless representations to controlled lossy approximations via rank truncation. This offers an alternative to particle-based compression strategies while retaining the possibility of high-accuracy solutions. Third, the tensor network formalism enables the governing equations to be expressed and advanced directly in compressed form, avoiding repeated reconstruction of the full phase-space state. When both the state and the relevant operators admit low-rank tensor network representations, linear maps can be applied efficiently through structured tensor contractions. A key example for the present work is the Fourier transform, which admits a highly compressible tensor representation~\cite{woolfe_scale_2014, chen_quantum_2023}, enabling spectral transforms to be applied directly in compressed form.

Motivated by these benefits, tensor network methods have emerged as an effective framework for the numerical simulation of dynamical systems. Although originally developed for compressed representations of quantum many-body states~\cite{vidal_efficient_2003, vidal_efficient_2004, schollwoeck_density-matrix_2011, orus_practical_2014}, these ideas have since been adapted into a broader class of \emph{quantum-inspired} methods for the simulation of complex classical systems. In this setting, notions such as entanglement and state compression are reinterpreted as low-rank structure and controllable approximation, providing a scalable algebraic toolbox for problems posed in exponentially large state spaces~\cite{brandejs_tensor_2026, manzini2026_tt_finite_difference}. Tensor-based techniques have consequently been applied to simulate a wide range of dynamical systems~\cite{khoromskij_tensor_2014, khoromskij_tensors-structured_2012}, including fluid dynamics~\cite{gourianov_quantum_2022, peddinti_quantum-inspired_2024, holscher_quantum-inspired_2025, gourianov_tensor_2025}, the Fokker--Planck equation~\cite{dolgov_fast_2012}, transport problems~\cite{truong2024_tt_neutron_transport}, and nonlinear wave equations~\cite{niedermeier_solving_2025}. In kinetic plasma simulation, tensor-based solvers have been proposed for the Vlasov--Poisson system using semi-Lagrangian schemes~\cite{kormann_semi-lagrangian_2015} as well as explicit time integrators combined with tensor-structured finite-difference operators~\cite{ye_quantum-inspired_2022}. Extensions to more general plasma models, including Vlasov--Maxwell systems, have also been investigated~\cite{allmann-rahn_parallel_2022, ye_quantized_2024}.

In this work, we introduce a spectral Vlasov--Poisson solver in which the phase-space distribution function $f$ is represented in the tensor train (TT) format. Our proposed method combines a fully spectral discretization with second-order Strang splitting~\cite{strang_construction_1968}, allowing the advection and acceleration substeps to be advanced efficiently in Fourier space while remaining within the TT formalism.
The advection substep is realized through the application of structured Fourier-space operators represented as matrix product operators, leading to efficient tensor contractions. The acceleration step, driven by the self-consistent electric field, is treated by reconstructing the updated tensor network state via tensor cross interpolation (TCI)~\cite{oseledets_tt-cross_2010, fernandez_learning_2025} from pointwise evaluations of the split-step update. A key feature of the proposed formulation is that low-rank Fourier-transform MPOs become explicit computational primitives within the time-evolution scheme. As a result, the split-step evolution can be expressed through a sequence of low-rank Fourier transforms, diagonal phase operators, and structured tensor contractions, yielding a fully spectral realization of the Vlasov--Poisson dynamics within the TT formalism. Numerical experiments show that the proposed solver reproduces the expected dynamics of standard benchmark problems, including Landau damping and the two-stream instability, while representing the phase-space distribution with up to several orders of magnitude fewer parameters than the corresponding full-grid representation.

This paper is organized as follows. Section~\ref{sec:background} introduces the required theoretical background, including a review of splitting methods for the integration of the Vlasov--Poisson equation and the \textit{quantics tensor train} encoding of the distribution function. In Section~\ref{sec:method}, we describe the spectral TT-based Vlasov--Poisson solver developed in this work. Sections~\ref{section:time_evol_in_TT_framework}--\ref{sec:compute_ef} detail the underlying theoretical formulation, while Sections~\ref{sec:complexity_analysis} and~\ref{sec:implementation_details} address computational complexity and implementation aspects. Numerical results obtained with the spectral TT-based solver are presented in Section~\ref{sec:results}, with Landau damping and the two-stream instability serving as benchmark problems. Section~\ref{sec:discussion} analyzes these results, highlighting both the strengths and limitations of the method and outlining directions for future work. Concluding remarks are given in Section~\ref{sec:conclusion}.

\section{Preliminaries}
\label{sec:background}

\subsection{Spectral splitting scheme of the Vlasov--Poisson equation}
\label{sec:vlasov_poisson}
The one-dimensional electron phase-space distribution function $f(x,v,t)$ in the presence of a static, uniform neutralizing background charge is governed by the 1D1V Vlasov--Poisson system
\begin{equation}
    \partial_t f + v\,\partial_x f - E(x)\,\partial_v f = 0 
    \label{eq:vlasov_eq}
\end{equation}
\begin{equation}
    \partial_x E = 1 - \int_{-\infty}^{\infty} f(x,v,t)\,dv
    \label{eq:poisson_eq}
\end{equation}
where the charge-to-mass ratio is set to $q/m = -1$ for convenience. 

The Vlasov--Poisson equation can be numerically integrated by employing a splitting scheme, which relies on separating the advection and acceleration of Eq.~\eqref{eq:vlasov_eq} into two distinct subproblems~\cite{strang_construction_1968, cheng_integration_1976, filbet_conservative_2001}. By defining the transport operators $\hat{\mathcal{A}}=-v\,\partial_x$ and $\hat{\mathcal{B}} = E(x)\, \partial_v$, the advection and acceleration subproblems can be formulated as
\begin{equation}
    \begin{cases}
        \partial_t f =\hat{\mathcal{A}}f , & \text{(advection)} \\
        \partial_t f =\hat{\mathcal{B}}f, & \text{(acceleration)}.
    \end{cases}
    \label{eq:splitting_subproblems}
\end{equation}
After Fourier transforming in the variable of differentiation, the time evolution generated by $\hat{\mathcal{A}}$ and $\hat{\mathcal{B}}$ reduces to diagonal phase shift operators in Fourier space. Denoting by $\hat{\mathcal{F}}_x$ and $\hat{\mathcal{F}}_v$ the Fourier transforms in space and velocity, the exact solutions of the two subproblems over a time increment $\Delta t$ can be written as
\begin{equation}
    \begin{cases}
        f(x, v, t + \Delta t)
        =
        \underbrace{
            \hat{\mathcal{F}}_x^{-1}
            \exp\!\left(- i k_x v \Delta t\right)
            \hat{\mathcal{F}}_x
        }_{\displaystyle \mathcal{S}_{\mathrm{adv}}(\Delta t)}
        \, f(x,v,t),
        & \text{(advection)},
        \\
        f(x, v, t + \Delta t)
        =
        \underbrace{
            \hat{\mathcal{F}}_v^{-1}
            \exp\!\left(i k_v E(x) \Delta t\right)
            \hat{\mathcal{F}}_v
        }_{\displaystyle \mathcal{S}_{\mathrm{acc}}(\Delta t; E)}
        \, f(x,v,t),
        & \text{(acceleration)}.
    \end{cases}
    \label{eq:solutions_fourier_space}
\end{equation}
Here $k_x = 2\pi n_x / (x_{\max}-x_{\min})$ and $k_v = 2\pi n_v / (v_{\max}-v_{\min})$ denote the discrete Fourier wavenumbers associated with the spatial and velocity domains, respectively, with $n_x$ and $n_v$ denoting the integer mode indices. The Fourier representation assumes periodic boundary conditions in the transformed variable. While periodicity in space is physical, periodicity in velocity is a numerical artifact introduced by truncating the velocity domain. Hence, we must choose a sufficiently large velocity interval such that the distribution function remains negligible near the velocity boundaries.

Time evolution according to the splitting scheme proceeds by alternately evolving the state $f(x,v,t)$ under the advection and acceleration generators. Using the update rules in Eq.~\eqref{eq:solutions_fourier_space}, we employ a second-order Strang splitting scheme~\cite{strang_construction_1968}. The solution is advanced from $t^n$ to $t^{n+1} = t^n + \Delta t$ by executing the following operations sequentially 
\begin{align}
    f^{(1)} 
    &=
    \hat{\mathcal{F}}_x^{-1}
       \exp\!\left(- ik_xv\Delta t/2\right)
    \hat{\mathcal{F}}_x \, f^{n},
    \label{eq:strang_half_adv}
    \\
    f^{(2)}
    &=
    \hat{\mathcal{F}}_v^{-1}
       \exp\!\left(ik_vE^{(1)}(x)\Delta t\right)
    \hat{\mathcal{F}}_v \, f^{(1)},
    \label{eq:strang_acc}
    \\
    f^{n+1}
    &=
    \hat{\mathcal{F}}_x^{-1}
       \exp\!\left(- ik_xv\Delta t/2\right)
    \hat{\mathcal{F}}_x \, f^{(2)}.
    \label{eq:strang_half_adv_2}
\end{align}
where $f^{(1)}$ and $f^{(2)}$ denote intermediate states, and the electric field $E^{(1)}(x)$ used in the acceleration step is computed self-consistently from $f^{(1)}$.

\begin{figure}
    \centering
    \includegraphics[width=1.0\columnwidth]{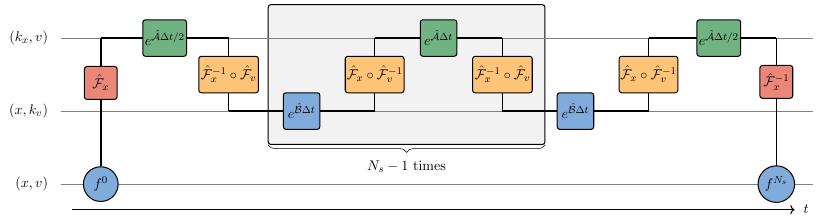}
    \caption{Workflow for the spectral Strang splitting integration scheme described in Eq.~\eqref{eq:strang_splitting_integration} over $N_s$ time steps. For both advection and acceleration, the solution is transformed to the Fourier domain of the variable of differentiation, after which the corresponding time propagator is applied. The highlighted block indicates the repeated inner loop applied $N_s-1$ times.}
    \label{fig:splitting_scheme_schematic}
\end{figure}

When this splitting scheme is applied repeatedly, the final half-advection of one time step and the initial half-advection of the subsequent step can be merged. Over $N_s$ time steps, the evolution may therefore be written compactly as
\begin{equation}
    f^{N_s}
    =
    \mathcal{S}_{\mathrm{adv}}\!\left(\tfrac{\Delta t}{2}\right)
    \mathcal{S}_{\mathrm{acc}}\!\left(\Delta t; E\right)
    \left[
        \mathcal{S}_{\mathrm{adv}}\!\left(\Delta t\right)
        \mathcal{S}_{\mathrm{acc}}\!\left(\Delta t; E\right)
    \right]^{N_s-1}
    \mathcal{S}_{\mathrm{adv}}\!\left(\tfrac{\Delta t}{2}\right)
    f^{0},
    \label{eq:strang_splitting_integration}
\end{equation}
where each $\mathcal S_{\mathrm{acc}}(\Delta t;E)$ is understood as the step-dependent acceleration map constructed from the self-consistent field at that step. This integration scheme is illustrated in Fig.~\ref{fig:splitting_scheme_schematic}.

\subsection{Compressed tensor representation of the phase-space distribution}
\label{sec:quantics}
It is possible to encode discretized functions, such as the phase-space distribution function, in a compressed tensor format using the so-called \textit{quantics tensor train} representation~\cite{khoromskij_odlog_2011, khoromskij_tensor_2014}.

We consider the distribution function $f(x,v)$ on a rectangular computational domain
$x \in [x_{\min},x_{\max})$, $v \in [v_{\min},v_{\max})$. For the spectral discretization, both coordinates are represented on periodic intervals. Periodicity in $x$ reflects the physical setting, whereas periodicity in $v$ is a numerical consequence of truncating the velocity domain. We let
\begin{equation}
i_x,i_v \in \{0,\dots,2^R-1\}
\end{equation}
denote the corresponding grid indices.
The associated grid coordinates are given by
\begin{equation}
x_{i_x}
=
x_{\min} + i_x \Delta x,
\qquad
v_{i_v}
=
v_{\min} + \left( i_v + \tfrac{1}{2} \right) \Delta v,
\end{equation}
with grid spacings
\begin{equation}
\Delta x
=
\frac{x_{\max}-x_{\min}}{2^R},
\qquad
\Delta v
=
\frac{v_{\max}-v_{\min}}{2^R}.
\end{equation}

A direct representation of the discretized distribution function
$f(x_{i_x},v_{i_v})$
as a $2^R \times 2^R$ array quickly becomes infeasible for large $R$, as both memory and computational costs scale exponentially.
The quantics approach circumvents this issue by expressing the grid indices in binary form,
\begin{equation}
i_x = \sum_{k=1}^R 2^{R-k} b^{(x)}_k, 
\qquad 
i_v = \sum_{k=1}^R 2^{R-k} b^{(v)}_k,
\end{equation}
with binary digits $b^{(x)}_k,b^{(v)}_k \in \{0,1\}$.
The sampled distribution function can then be reinterpreted as an order-$N$ tensor, with $N=2R$, given by
\begin{equation}
F_{b^{(x)}_1 b^{(v)}_1 \dots b^{(x)}_R b^{(v)}_R}
:=
f(x_{i_x},v_{i_v}),
\end{equation}
where the tensor indices correspond directly to the binary digits of the spatial and velocity grid indices.

\begin{figure}[h]
    \centering
    \includegraphics[width=.9\columnwidth]{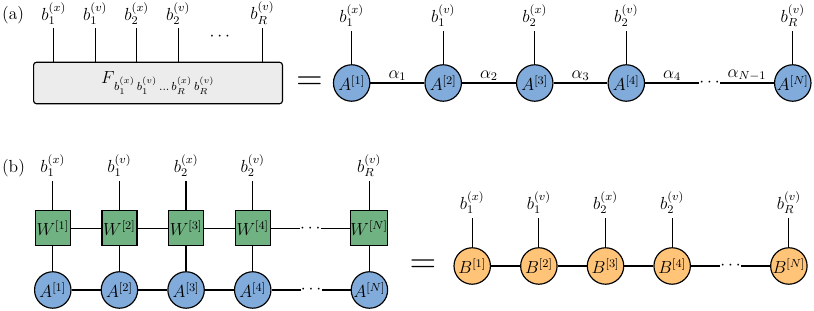}
    \caption{\textbf{(a)} The discretized distribution function $f(x_{i_x}, v_{i_v})$ is encoded as an order-$2R$ tensor using the quantics representation and represented in tensor train (TT) format as a chain of local tensors connected by auxiliary bond indices. In a tensor network diagram, nodes denote tensors, while edges connecting two nodes indicate summation (contraction) over shared indices. Edges that terminate without connecting to another node represent free (site) indices of the tensor. \textbf{(b)} Application of a matrix product operator (MPO) to the TT representation of $f$, yielding a new TT with the same physical structure. The action of the MPO is performed by contracting the site indices of the TT with the corresponding input indices of the MPO at each site. Approximation is introduced only if the resulting TT is subsequently recompressed by bond-dimension truncation.
    }
    \label{fig:TT_quantics}
\end{figure}

This high-order tensor can be represented exactly in \emph{tensor train} (TT) format as a product of lower-order tensors connected through a set of auxiliary indices,
\begin{equation}
    F_{\,b^{(x)}_1\, b^{(v)}_1\, \dots\, b^{(x)}_R\, b^{(v)}_R}
    =
    \sum_{\alpha_1,\ldots,\alpha_{2R-1}}
    A^{(1)}_{\,b^{(x)}_1\alpha_1}\,
    A^{(2)}_{\,\alpha_1 b^{(v)}_1 \alpha_2}\,
    \ldots\,
    A^{(2R-1)}_{\,\alpha_{2R-2} b^{(x)}_R \alpha_{2R-1}}\,
    A^{(2R)}_{\,\alpha_{2R-1} b^{(v)}_R}
\end{equation}
effectively factorizing the full tensor
$F_{\,b^{(x)}_1\, b^{(v)}_1\, \dots\, b^{(x)}_R\, b^{(v)}_R}$
in terms of the set of smaller local tensors $\{A^{(k)}\}_{k=1}^{2R}$ (see Fig.~\ref{fig:TT_quantics}(a)).
Here, each local tensor $A^{(k)}$ carries one binary \emph{site index} and two auxiliary \emph{bond indices}. The corresponding bond dimensions determine the complexity of the TT representation. In the worst case, an exact TT representation may require bond dimensions that grow exponentially with the number of sites. However, many physically relevant tensors admit accurate low-rank approximations with bounded maximum bond dimension $\chi$. In such cases, the TT representation requires storing only $\mathcal{O}(N\chi^2)$ entries instead of the $\mathcal{O}(2^N)$ entries of the full tensor, thereby yielding exponential compression.

Crucially, the tensor cross interpolation (TCI) algorithm allows for the direct construction of a compressed TT approximation of the discretized distribution function $f(x_{i_x},v_{i_v})$, without having to store the full tensor $F_{\,b^{(x)}_1\, b^{(v)}_1\, \dots\, b^{(x)}_R\, b^{(v)}_R}$~\cite{oseledets_tensor-train_2011, dolgov_parallel_2020, fernandez_learning_2025}.
Instead, TCI assumes that the discretized target function $f(x_{i_x},v_{i_v})$ can be evaluated at arbitrary grid points on demand, without assembling the full tensor, and constructs an approximate TT representation by querying only $\mathcal{O}(Nd\chi^2)$ values, where $d=2$ denotes the local site dimension. 

The TCI algorithm constructs the TT approximation through a sweeping procedure across the TT chain. During each sweep, either individual local tensors or neighboring tensor pairs are updated so as to reduce the interpolation error. These local updates are determined from adaptively selected interpolation points, referred to as \textit{pivots}. The sweeps proceed back and forth across the chain until the target error tolerance is reached or a prescribed maximum number of sweeps has been performed. Unlike the full tensor $F_{\,b^{(x)}_1\, b^{(v)}_1\, \dots\, b^{(x)}_R\, b^{(v)}_R}$, the TT representation provided by TCI is generally only approximate and yields a compressed low-rank representation whose bond dimensions are determined adaptively by the compressibility of the target tensor and the desired error tolerance. For a detailed discussion of the TCI algorithm, we refer to the overview provided in~\cite{fernandez_learning_2025}.

Although the ordering of the binary indices within the TT is, in principle, arbitrary, the choice of ordering can have a significant impact on the compressibility of the approximate TT state~\cite{ye_quantum-inspired_2022}. In this work, we employ an \emph{interleaved} ordering in which bits corresponding to position and velocity alternate along the TT chain. This choice places bits encoding comparable physical scales at neighboring sites, allowing the TT representation to efficiently capture correlations between position and velocity across scales.

A key advantage of the TT formalism is that linear operators acting on TT-encoded functions can themselves be represented in compressed tensor-network form as matrix product operators (MPOs), enabling their efficient application while remaining entirely within the tensor framework. Operators such as Fourier transforms can be applied efficiently through tensor contractions (see Fig.~\ref{fig:TT_quantics}(b)), a property that is central to the spectral solver developed in this work.

\section{Spectral tensor train Vlasov--Poisson solver}
\label{sec:method}

\subsection{Time evolution within the tensor train framework}
\label{section:time_evol_in_TT_framework}

Starting from a quantics TT representation of the discretized distribution function $f^n(x_{i_x},v_{i_v})$
in the physical $(x,v)$ basis, the solver advances the state to $f^{n+1}(x_{i_x},v_{i_v})$ through a second-order Strang splitting scheme consisting of alternating advection and acceleration steps.

The advection propagator is diagonal in Fourier space with respect to the spatial coordinate.
Therefore, the TT state representation is first transformed from the physical $(x,v)$ basis to the mixed Fourier basis $(k_x,v)$ through application of the spatial Fourier-transform MPO. The advection propagator is subsequently applied through a second TT--MPO contraction. Finally, application of the inverse spatial Fourier MPO then returns the state to the physical $(x,v)$ basis. We discuss the construction of the MPO representations of the Fourier transform and advection propagator in Section~\ref{section:mpo_construction}.

Exact application of advection and Fourier MPOs to TT-represented states increases the bond dimension of the resulting TT. To maintain a compressed representation throughout time evolution, each TT--MPO contraction is therefore followed by a truncation step in which the bond dimensions are reduced through a sequence of local singular value decompositions (SVDs). At each bond, the smallest singular values $\{\sigma_n\}$ associated with the corresponding bipartition of the TT are discarded such that the truncation error,
\begin{equation}
    \frac{\sum_{n \in \mathrm{discarded}} \sigma_n^2}
    {\sum_n \sigma_n^2}
    < \epsilon,
\end{equation}
remains below a prescribed threshold $\epsilon$. This procedure produces a compressed approximate TT representation while controlling the growth of the bond dimensions throughout the time evolution.

To execute the acceleration step, the state is first transformed to the mixed basis $(x,k_v)$ through application of the velocity Fourier-transform MPO. The electric field $E(x_{i_x})$ is then computed from the TT representation of the distribution function in the mixed basis, as described in Section~\ref{sec:compute_ef}. Rather than constructing an acceleration MPO that represents the phase factor
$\exp(iE(x_{i_x})k_{i_v}\Delta t)$ induced by acceleration, we directly approximate
the result of the acceleration step in the Fourier basis using TCI. Specifically, TCI is used to approximate the function
\begin{equation}
\hat{\Theta}(n_v)\,
\exp(i E(x_{i_x}) k_{i_v} \Delta t)\,
f(x_{i_x}, k_{i_v}),
\label{eq:TCI_fit_kernel}
\end{equation}
where $\hat{\Theta}$ denotes the spectral filter introduced below. This procedure directly produces a compressed TT approximation of the accelerated state in the $(x,k_v)$ basis, thereby circumventing the need to execute an additional TT--MPO contraction. Application of the inverse velocity Fourier MPO subsequently returns the state to the physical $(x,v)$ basis.

To suppress high-frequency modes and improve compressibility during the split-step updates, we introduce the Fourier-space spectral filter $\hat{\Theta}(n)$, defined as 
\begin{equation}
\hat{\Theta}(n)
=
\frac{1}{\exp\!\bigl((|n|-n_\text{max})\beta\bigr) + 1},
\end{equation} 
where $\beta$ controls the sharpness of the cutoff and $n_\text{max}$ sets the center of the spectral cutoff and therefore approximates the largest effectively retained mode index. The filter is applied in the transformed coordinate in both the advection and acceleration substeps. In the advection step, it reduces the complexity of the compressed MPO representation of the phase factor, whereas in the acceleration step it improves the compressibility of the updated state reconstructed by TCI.

\begin{figure}[h]
    \centering
    \includegraphics[width=.9\columnwidth]{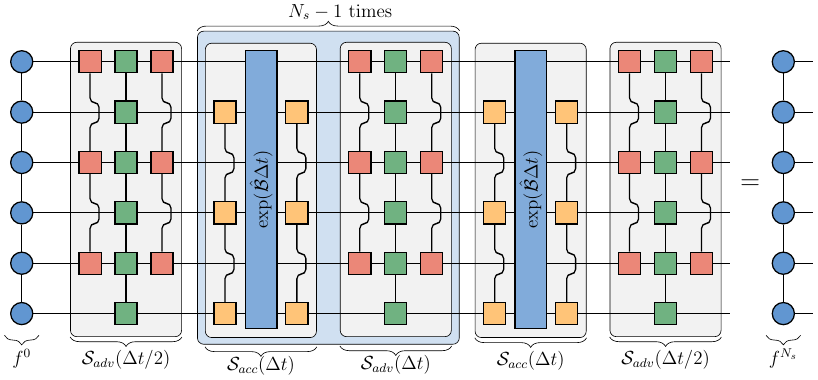}
    \caption{Time evolution of the TT representation of the initial condition $f^0$, following the sequence dictated by the second-order Strang splitting scheme. Advection steps are implemented via TT--MPO contractions, whereas the acceleration steps are realized using tensor cross interpolation (TCI). Fourier transforms in position (red) and velocity (yellow) act only on the corresponding TT site registers, whereas the phase shifts associated with advection and acceleration (green and blue) act on the full position--velocity register.}
    \label{fig:algorithm_schematic}
\end{figure}

A distinguishing feature of the present formulation is that low-rank Fourier-transform MPOs enter directly as computational primitives of the time-evolution algorithm. Consequently, spectral information remains directly accessible throughout the computation, naturally enabling extensions based on spectral filtering, adaptive mode truncation, and other frequency-space operations. We provide a schematic overview of the time evolution scheme in Fig.~\ref{fig:algorithm_schematic}.

\subsection{Charge and momentum conservation}
We ensure charge and momentum conservation by using a projection method~\cite{hairer_conservation_2006}. On the discretized grid, charge and momentum are defined by the quantities
\begin{equation}
    Q(f) = \Delta x \Delta v \sum_{i_x, i_v} f(x_{i_x}, v_{i_v}), \quad P(f) = \Delta x \Delta v \sum_{i_x, i_v} v_{i_v} f(x_{i_x}, v_{i_v})
    \label{eq:charge-and-momentum}
\end{equation}
Given the target values $Q_0 = Q(f^0)$ and $P_0=P(f^0)$, we define the conservation defects
\begin{equation}
    \delta Q =  Q_0 - Q(\tilde{f}), \quad \delta P = P_0 - P(\tilde{f})
\end{equation}
where $\tilde{f}$ denotes the uncorrected state after a time step. Seeking the minimum-norm correction subject to the conservation constraints implies that the correction lies in the span of the corresponding constraint gradients. For charge and momentum conservation, these gradients are proportional to $1$ and $v$, hence giving a corrected state of the form
\begin{equation}
    f = \tilde{f} + \lambda_0 1 + \lambda_1 v.
\end{equation}
For a velocity domain symmetric about zero, the midpoint-centered velocity grid satisfies $\sum_{i_v} v_{i_v}=0$ exactly, which implies that the functions $1$ and $v$ are orthogonal on the discretized domain. In this case, the correction coefficients are given by
\begin{equation}
    \lambda_0 = \frac{\delta Q}{\Delta x \Delta v N_x N_v}, \quad \lambda_1 = \frac{\delta P}{\Delta x \Delta v N_x\sum_{i_v} v_{i_v}^2}
\end{equation}
where $N_x$ and $N_v$ denote the number of grid points along the position and velocity domains.

The projection method is suitable for the tensor representation, as the functions $1$ and $v$ admit exact low-rank representations with bond dimensions $\chi = 1$ and $\chi = 2$, respectively. Furthermore, exact TT addition increases the bond dimensions additively, with the resulting TT having bond dimensions equal to the sums of the corresponding bond dimensions of the original TTs. The corrected state $f$ therefore has bond dimensions that are at most 3 larger than those of $\tilde f$. To ensure sustained conservation, the correction is applied to the state at the end of every time step.

\subsection{Construction of MPO propagators}
\label{section:mpo_construction}
The discrete Fourier transforms in position and velocity required for the spectral splitting scheme, together with their inverses, are implemented using the analytic MPO construction introduced by Chen et al.~\cite{chen_quantum_2023}, which achieves machine precision with a bond dimension of only $\chi \approx 11$. Throughout this work, we employ a unitary normalization of the discrete Fourier transform, with both the forward and inverse transforms carrying a prefactor of $1/\sqrt{2^R}$. The Fourier transforms act only on the subset of $R$ sites of the TT associated with either the position or velocity register. To embed the Fourier MPOs into the full position-velocity register, we insert identity operators on all sites corresponding to the complementary coordinate. An important aspect of the compressed MPO representation of the Fourier transform is that it reverses the order of the quantics bit register. As a result, subsequent operations in spectral Fourier space are carried out with respect to this reversed bit ordering.

To construct the MPO representation of the advection propagator $\exp(\hat{\mathcal A}\,\Delta t)$, we exploit the fact that $\hat{\mathcal A}=-v\,\partial_x$ is diagonal in the Fourier representation with respect to the spatial coordinate. Conjugation with the spatial Fourier transform yields
\begin{equation}
\exp(\Delta t\,\hat{\mathcal A})
=
\hat{\mathcal F}_x^{-1}
\exp(- i k_x v \Delta t)
\hat{\mathcal F}_x ,
\end{equation}
such that time evolution under $\hat{\mathcal A}$ reduces to pointwise multiplication by the scalar phase factor $g(k_x,v)=\exp(- i k_x v \Delta t)$ in $(k_x,v)$-space.

To represent this phase factor in quantics TT form, we evaluate $g(k_x,v)$ on the discretized quantics grid associated with the binary indices
$\tilde b^{(x)}_1,\ldots,\tilde b^{(x)}_R$ and
$\tilde b^{(v)}_1,\ldots,\tilde b^{(v)}_R$.
This defines the order-$2R$ tensor
\begin{equation}
G_{\tilde b^{(x)}_1 \tilde b^{(v)}_1 \dots \tilde b^{(x)}_R \tilde b^{(v)}_R}
:=
g(k_x,v),
\end{equation}
where $(k_x,v)$ are the grid coordinates corresponding to the binary multi-index. The tensor $G$ is then approximated in TT form using TCI,
\begin{equation}
\mathcal G_{\tilde b^{(x)}_1 \tilde b^{(v)}_1 \dots \tilde b^{(x)}_R \tilde b^{(v)}_R}
\approx
\hat{\Theta}(n_x) G_{\tilde b^{(x)}_1 \tilde b^{(v)}_1 \dots \tilde b^{(x)}_R \tilde b^{(v)}_R},
\end{equation}
where $\hat{\Theta}$ denotes the spectral filtering operator introduced in Section~\ref{section:time_evol_in_TT_framework}.

While $\mathcal G$ approximates the scalar phase factor in TT form, applying the phase shift to a TT-represented distribution function requires an operator representation. We therefore promote the TT representation of $\mathcal G$ to a diagonal MPO $\mathcal{M}$ through contraction with three-index Kronecker deltas, thereby enforcing the equality of input and output indices at each site:
\begin{equation}
\mathcal{M}_{\,
b^{(x)}_1\, b^{(v)}_1\, \dots\, b^{(x)}_R\, b^{(v)}_R;\,
b^{(x)\prime}_1\, b^{(v)\prime}_1\, \dots\, b^{(x)\prime}_R\, b^{(v)\prime}_R
}
=
\sum_{\substack{\tilde{b}_1^{(x)}, \ldots, \tilde{b}_R^{(x)} \\ \tilde{b}_1^{(v)}, \ldots, \tilde{b}_R^{(v)}}}
\mathcal{G}_{\,\tilde b^{(x)}_1\, \tilde b^{(v)}_1\, \dots\, \tilde b^{(x)}_R\, \tilde b^{(v)}_R}
\prod_{r=1}^R
\delta_{\,b^{(x)}_r\, b^{(x)\prime}_r\, \tilde b^{(x)}_r}\,
\delta_{\,b^{(v)}_r\, b^{(v)\prime}_r\, \tilde b^{(v)}_r}.
\label{eq:phase_shift_MPO}
\end{equation}
Contracting the MPO $\mathcal M$ with a TT has the desired effect of multiplying by $\exp(- i k_{i_x} v_{i_v} \Delta t)$ pointwise on the quantics grid. In Eq.~\eqref{eq:phase_shift_MPO}, we associate the input (output) indices with the unprimed (primed) indices. 

We emphasize that unlike finite-difference approaches, which approximate the action of the propagator through a finite-step integration scheme, the present formulation constructs a compressed approximation of the advection propagator itself. As a result, the accuracy of the advection operator is governed by the spectral resolution and TCI reconstruction tolerance rather than by the local truncation error associated with a finite-step integration method. Furthermore, the advection MPO is time-independent and needs to be constructed only once.

\subsection{Computing the electric field}
\label{sec:compute_ef}

\begin{figure}[h]
    \centering
    \includegraphics[width=.9\columnwidth]{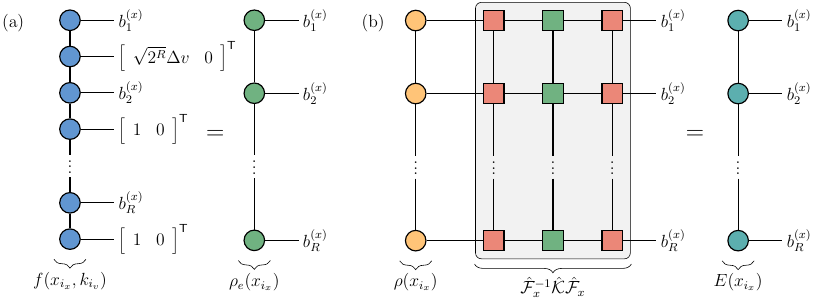}
    \caption{\textbf{(a)} The electron density $\rho_e$ is computed by contracting the tensors in the velocity register by $\begin{bmatrix} 1 & 0 \end{bmatrix}^{\mathsf T}$, effectively selecting the zero-mode. One site must be contracted by $\begin{bmatrix} \sqrt{2^R}\,\Delta v & 0 \end{bmatrix}^{\mathsf T}$, to account for the spacing of the velocity grid and to compensate for the prefactor of the unitary Fourier transform. \textbf{(b)} The electric field is obtained by multiplying the TT representation of the charge density $\rho(x_{i_x})$ by a sequence of three MPOs, corresponding to Eq.~\eqref{eq:poisson_fourier_kernel}.}
    \label{fig:poisson_solver}
\end{figure}

To carry out the acceleration step in Eq.~\eqref{eq:strang_acc} of the splitting scheme, the electric field must be computed from the TT representation of the state~$f$. As described below, this can be accomplished while remaining entirely within the tensor framework.

We first compute the discretized electron density $\rho_e(x_{i_x})$ by summing the distribution function over the velocity domain
\begin{equation}
    \rho_e(x_{i_x})
    =
    \Delta v\sum_{i_v} f(x_{i_x}, v_{i_v}).
    \label{eq:electron_density_def}
\end{equation}
Assuming that the velocity register of the quantics TT representing $f$ is given in the Fourier basis, the summation corresponds to selecting the zero Fourier mode. Because a unitary discrete Fourier transform is used, the zero mode equals the velocity sum divided by $\sqrt{2^R}$. The density extraction therefore amounts to contracting each velocity site with the vector $\begin{bmatrix} 1 & 0 \end{bmatrix}^{\mathsf T}$, except for a single site that is contracted with $\begin{bmatrix} \sqrt{2^R}\,\Delta v & 0 \end{bmatrix}^{\mathsf T}$. The resulting object is a rank-$R$ TT defined solely over the spatial register, representing the spatial electron density $\rho_e(x_{i_x})$. A schematic of the tensor contraction is shown in Fig.~\ref{fig:poisson_solver}(a).

Including a uniform ion background with unit density, the total charge density is then given by
\begin{equation}
    \rho(x_{i_x}) = 1 - \rho_e(x_{i_x}).
\end{equation}
The electric field $E(x_{i_x})$ is obtained by solving the Poisson equation~\eqref{eq:poisson_eq} in Fourier space
\begin{equation}
E(x_{i_x})
=
\hat{\mathcal{F}}_x^{-1}\! \hat{\mathcal{K}}(k_{i_x})\,\hat{\mathcal{F}}_x \rho(x_{i_x}),
\qquad
\hat{\mathcal{K}}(k_{i_x})=
\begin{cases}
-\dfrac{i}{k_{i_x}}, & k_{i_x} \neq 0,\\[6pt]
0, & k_{i_x} = 0,
\end{cases}
\label{eq:poisson_fourier_kernel}
\end{equation}
where the zero mode $k_{i_x}=0$ is set to zero to enforce a zero-mean electric field. The Fourier-space multiplier $\hat{\mathcal{K}}(k_{i_x})$ is represented as an MPO constructed using TCI, following the same methodology employed for the advection propagator described in Section~\ref{section:mpo_construction}. The quantics TT representation of the electric field is then obtained through a sequence of TT--MPO contractions, as illustrated in Fig.~\ref{fig:poisson_solver}(b). 

\subsection{Complexity analysis}
\label{sec:complexity_analysis}
Next, we summarize the scaling of the steps of the TT-based time-splitting scheme.

Constructing the analytic Fourier MPOs with maximum bond dimension $\chi$ scales as $\mathcal{O}(R^2 \chi^3)$~\cite{chen_quantum_2023}. This cost is incurred only once for a given quantics grid and is therefore a preprocessing step that does not add to the per-time-step runtime of the algorithm.

Fitting the MPO that represents the phase shift induced by advection using TCI scales as $\mathcal{O}(N d \chi^3)$~\cite{fernandez_learning_2025, dolgov_parallel_2020}, where $\chi$ denotes the maximum bond dimension of the resulting tensor and $d=2$ is the dimension of the binary site indices. As in the case of the Fourier MPOs, the advection MPO depends only on the chosen quantics grid and therefore needs to be constructed only once. The same $\mathcal{O}(N d \chi^3)$ scaling applies to the TCI fit used to obtain the evolved state after the acceleration step in Eq.~\eqref{eq:TCI_fit_kernel}, assuming constant-cost access to pointwise tensor entries. In the present implementation, this assumption is generally not satisfied, since each query requires separate TT contractions to evaluate $E(x_{i_x})$ and $f(x_{i_x},k_{i_v})$. Therefore, the stated scaling describes the interpolation step itself rather than the full cost of the acceleration update. As this fit is performed once every time step, it constitutes a recurring contribution to the total computational cost.

The TT--MPO contractions required to perform the Fourier transforms and advection time evolution on the state $f$ scale as $\mathcal{O}(Nd^2\chi_\text{TT}^2\chi_\text{MPO}^2)$, where $\chi_\text{TT}$ and $\chi_\text{MPO}$ denote the maximum bond dimensions of the TT and MPO, respectively~\cite{schollwoeck_density-matrix_2011, paeckel_time-evolution_2019}. Furthermore, the SVD-based truncation sweep used to truncate the bond dimensions after each TT--MPO contraction has a scaling of $\mathcal{O}(Nd^2\chi_\text{TT}^3\chi_\text{MPO})$.

We emphasize that all tensor-network operations contributing to the per-step runtime scale linearly with respect to the number of bits $N=2R$ used in the quantics representation. Under the additional assumption of constant-cost tensor-entry queries in the TCI procedure, the time-stepping complexity is likewise linear in $N$. Hence, provided the bond dimensions of the state $f$ and the MPO operations remain bounded, we expect an exponential speed-up over conventional methods operating on the full $2^R\times 2^R$ grid.

\subsection{Implementation details}
\label{sec:implementation_details}
We execute numerical simulations on a quantics grid using $R =10$ bits for both the spatial and velocity dimensions, corresponding to $2^R$ grid points per dimension. In total, we therefore operate on a position-velocity grid with $2^{20} \approx 10^6$ grid points. Time integration is carried out using a fixed time step $\Delta t = 0.1$. Unless otherwise stated, the default value of the spectral filtering cutoff is set to $n_\text{max}=2^6$, with the cutoff sharpness fixed at $\beta = 2$.

For the acceleration steps implemented with TCI, we use the standard two-site TCI algorithm, which optimizes neighboring pairs of TT cores during sweeps across the full train. Pivot selection uses the \textit{full} strategy, meaning that all entries of the local two-site tensors are evaluated when choosing pivots.

For each acceleration step, the TCI optimization is initialized from the TCI representation obtained at the previous time step rather than from a random initial guess. In particular, the previously selected interpolation pivots and the associated bond-dimension structure are reused, yielding a warm start of the TCI optimization. This is motivated by the fact that the state changes only incrementally between consecutive time steps.

The TCI stopping tolerance is fixed at $\tau = 10^{-8}$. The TCI tolerance corresponds to a sampled proxy for the relative $L^\infty$ approximation error, evaluated on the adaptively queried interpolation points rather than on the full tensor domain. The TCI optimization is allowed to perform at most 20 back-and-forth two-site sweep iterations and terminates earlier if the target tolerance is reached. Upon termination, a final one-site refinement sweep is performed.

For the SVD-based truncation executed after each MPO contraction, the truncation threshold $\epsilon$ is varied between simulations in order to investigate the effect of TT compression on the numerical accuracy and stability of the solver. For the TT representation of the electric field, a fixed maximum bond dimension of $\chi_E=32$ is imposed. 

Crucially, no explicit upper bound is imposed on the TT bond dimensions throughout time evolution. Rank control is introduced only by the TCI tolerance $\tau$ and the threshold $\epsilon$ of the SVD-based truncation step. In this way, the bonds are allowed to grow adaptively according to the complexity of the represented distribution function throughout time evolution.

A summary of the numerical and algorithmic parameters is provided in Table~\ref{tab:simulation_parameters}. Simulations are implemented in \texttt{Julia} using \texttt{ITensors.jl}~\cite{fishman_itensor_2022} and \texttt{TCI.jl}~\cite{fernandez_learning_2025}, and are performed in double-precision arithmetic (\texttt{Float64}). TT/MPO contractions are executed on an NVIDIA RTX 5070 GPU, while TCI logic and tensor-entry queries are performed on an AMD Ryzen 7 9700X CPU.

\begin{table}[t]
    \centering
    \setlength{\tabcolsep}{25pt} 
    \caption{Summary of numerical and algorithmic parameters used in generating the numerical results.}
    \label{tab:simulation_parameters}
    \renewcommand{\arraystretch}{1.0}
    \begin{tabular}{l c}
        \toprule
        \textbf{Parameter} & \textbf{Value} \\
        \midrule
        Quantics bits per dimension ($x$, $v$) & $R = 10$ \\
        Grid points per dimension & $2^R = 1024$ \\
        Total phase-space grid size & $2^{20} \approx 10^6$ \\
        Time step & $\Delta t = 0.1$ \\
        TCI tolerance & $\tau = 10^{-8}$ \\
        TT truncation cutoff & $\epsilon$ (varied) \\
        Electric field max. bond dimension & $\chi_E = 32$ \\
        \bottomrule
    \end{tabular}
\end{table}

\section{Numerical results}
\label{sec:results}
In this section, we investigate the accuracy, stability, and computational behavior of the proposed TT-based spectral Vlasov--Poisson solver. The algorithm is assessed using Landau damping and the two-stream instability as complementary benchmarks. The TTs corresponding to the initial conditions are obtained via TCI. Unless otherwise stated, all simulations are performed using the fixed parameters specified in Table~\ref{tab:simulation_parameters}.

In evaluating the numerical protocol, we monitor the charge $Q$, momentum $P$, and total energy $\mathcal{E} = \mathcal{E}_\text{field} + \mathcal{E}_\text{kin}$. The discrete definitions of $Q$ and $P$ are given in Eq.~\eqref{eq:charge-and-momentum}, while the field and kinetic energy are computed as 
\begin{equation}
    \mathcal{E}_\text{field}= \frac{\Delta x}{2} \sum_{i_x} \lvert E(x_{i_x}) \rvert^2, \quad
    \mathcal{E}_\text{kin} = \frac{\Delta x\Delta v}{2} \sum_{i_x,i_v} v_{i_v}^2 f(x_{i_x}, v_{i_v})
\end{equation}
with $\Delta x$ and $\Delta v$ denoting the position and velocity grid point spacing. All observables are evaluated efficiently via tensor contractions within the TT formalism. We also monitor the energy contained in the first Fourier mode pair of the electric field, defined as
\begin{equation}
    \mathcal{E}_\text{field}^{(1)}
    =
    \frac{\Delta x}{2}
    \left(
        |\hat{E}_{1}|^2
        +
        |\hat{E}_{-1}|^2
    \right),
\end{equation}
where $\hat{E}_{\pm 1}$ denote the discrete Fourier coefficients of the electric field corresponding to the mode indices $n=\pm 1$.

\subsection{Physics validation}
We first validate the solver by comparing the decay or growth rate of the first electric-field mode against analytical predictions in the linear regime. For both Landau damping and the two-stream instability, we execute the validation simulations using a strict singular value truncation threshold of $\epsilon = 10^{-10}$. As a reference solver, we run the spectral Strang-splitting scheme on a full grid representation using the same grid resolution and time step, and without imposing any frequency cutoff in the advection and acceleration phase shifts.

\paragraph{Linear Landau damping}
We simulate linear Landau damping on the phase-space domain $x\in[-2\pi,2\pi]$ and $v\in[-6,6]$. The initial distribution is set to a Maxwellian with a first Fourier-mode spatial perturbation,
\begin{equation}
f(x,v,0)= \frac{1}{\sqrt{2\pi}\,v_\text{th}}
\exp\!\left(-\frac{v^2}{2v_\text{th}^2}\right)
\left(1 + A \cos\left(\frac{2\pi x}{x_\text{max}- x_\text{min}}\right)\right),
\label{eq:ic_landau_damping}
\end{equation}
with thermal velocity $v_\text{th}=1.0$ and perturbation amplitude $A=0.01$. For this standard benchmark, linear theory predicts exponential decay of the amplitude of the first Fourier mode of the electric field at a rate $\gamma_t = -0.151\,\omega_{pe}$, corresponding to decay of its energy at rate $2\gamma_t$. We use this decay rate as a reference value when assessing the numerical results.

\begin{figure}
    \centering
    \includegraphics[width=.82\textwidth]{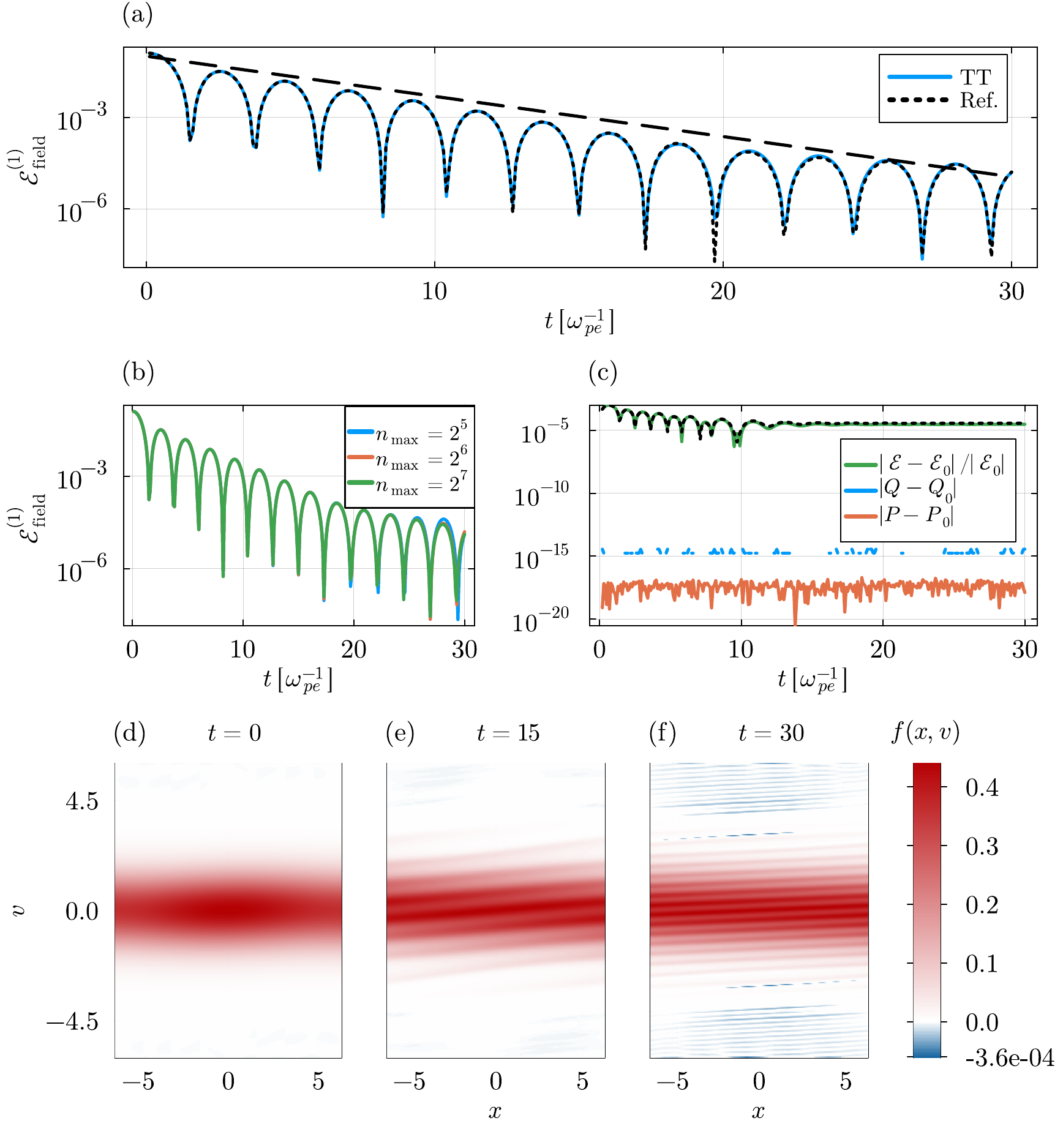}
    \caption{TT-based simulation of Landau damping, with initial condition set according to Eq.~\eqref{eq:ic_landau_damping}.
    \textbf{(a)} Energy of the first mode of the electric field $\mathcal{E}_\text{field}^{(1)}$. The TT-based simulation is in excellent agreement with the full-grid reference (dotted line) and reproduces the analytically expected damping rate (dashed line).
    \textbf{(b)} Damping of the energy of the first mode of the electric field $\mathcal{E}_\text{field}^{(1)}$ obtained for different frequency cutoffs $n_\text{max}=2^5, 2^6, 2^7$, all of which result in near-identical damping behavior.
    \textbf{(c)} Evolution of charge deviation $\lvert Q-Q_0 \rvert$, momentum deviation $\lvert P - P_0 \rvert$, and relative energy deviation $|\mathcal{E}-\mathcal{E}_0|/|\mathcal{E}_0|$ during Landau damping. The dotted curve shows the energy-deviation trajectory of the full-grid reference. Apparent gaps in the charge-deviation curve correspond to time steps where the error is exactly zero and therefore cannot be displayed on the logarithmic y-axis.
    \textbf{(d--f)} Snapshots of the distribution function $f$, showing gradual filamentation along the velocity domain.}
    \label{fig:physical_validation_landau}
\end{figure}

The simulation results for Landau damping are presented in Fig.~\ref{fig:physical_validation_landau}. As shown in Fig.~\ref{fig:physical_validation_landau}(a), the temporal decay of the energy associated with the first Fourier mode of the electric field $\mathcal{E}_\text{field}^{(1)}$ is in good agreement with the analytical prediction, demonstrating that the expected Landau damping rate is accurately captured. Furthermore, the TT-based solver produces an energy damping trajectory that is indistinguishable from that of the full-grid reference solver.

To isolate any potential accuracy loss emerging from the spectral filtering operator $\hat{\Theta}$, we run repeated Landau damping simulations while sweeping the frequency index cutoff $n_\text{max}$, and present the results in Fig.~\ref{fig:physical_validation_landau}(b). We find that the frequency cutoff values $n_\text{max} = 2^5, 2^6, 2^7$ all result in near-identical damping behavior throughout the full simulation. Therefore, imposing the default cutoff of $n_\text{max}=2^6$ does not appear to impose any significant accuracy loss for the simulation conditions considered here.

We study the conservation of charge, momentum, and energy in Fig.~\ref{fig:physical_validation_landau}(c). For charge and momentum, we observe maximum deviations from the initial values on the order of $|Q-Q_0| \sim 10^{-15}$ and $|P-P_0| \sim 10^{-17}$, confirming that the projection-based corrections provide conservation of charge and momentum to machine precision. Concerning the relative energy deviation $|\mathcal{E}-\mathcal{E}_0|/|\mathcal{E}_0|$, the error is larger than that of the charge and momentum, as no direct measures were taken to ensure energy conservation. Despite this, however, the energy deviation stabilizes at a level of $|\mathcal{E}-\mathcal{E}_0|/|\mathcal{E}_0| \approx 5 \cdot 10^{-5}$ after the initial fluctuations. Notably, the relative energy error obtained by the full-grid reference remains nearly identical to that of the TT-based solver throughout the full simulation, suggesting that conservation errors emerge from the integration method itself, rather than from the compressed state representation.

Finally, snapshots of the distribution function $f$ at selected times are shown in Fig.~\ref{fig:physical_validation_landau}(d)--(f), clearly illustrating the progressive filamentation characteristic of Landau damping. As time evolution progresses, small negative numerical artifacts develop, with the minimum value reaching $f_{\min} \approx -3.6 \times 10^{-4}$ at the final step.

\paragraph{Two-stream instability}
We consider the two-stream instability on the domain
\( x \in [-\pi/3.06, \pi/3.06] \), \( v \in [-0.6, 0.6] \). The initial distribution consists of two counter-propagating Maxwellian beams with a superimposed first Fourier-mode spatial perturbation,
\begin{equation}
f(x,v,0) = (f_+(x,v) + f_-(x,v))\left(1 + A \cos\!\left(\frac{2\pi x}{x_\text{max}-x_\text{min}}\right)\right)
\label{eq:ic_two_stream}     
\end{equation}
with the beams defined as
\begin{equation}
f_{\pm}(x,v)= \frac{1}{2\sqrt{2\pi}v_\text{th}}
\exp\!\left(-\frac{(v\pm v_0)^2}{2v_\text{th}^2}\right).
\label{eq:ic_two_stream_2}
\end{equation}
In Eqs.~\eqref{eq:ic_two_stream} and~\eqref{eq:ic_two_stream_2}, the thermal velocity is \( v_\text{th} = 0.01 \), the perturbation amplitude is \( A = 0.01 \), and the bulk beam velocity is \( v_0 = 0.2 \). In the linear regime, this configuration is unstable and exhibits exponential growth of the electric field amplitude. For the parameters used here, linear theory predicts a growth rate $\gamma_t = 0.354\omega_{pe}$ of the amplitude of the first electric field mode, corresponding to a modal-energy growth rate of $2\gamma_t$.

\begin{figure}
    \centering
    \includegraphics[width=.82\textwidth]{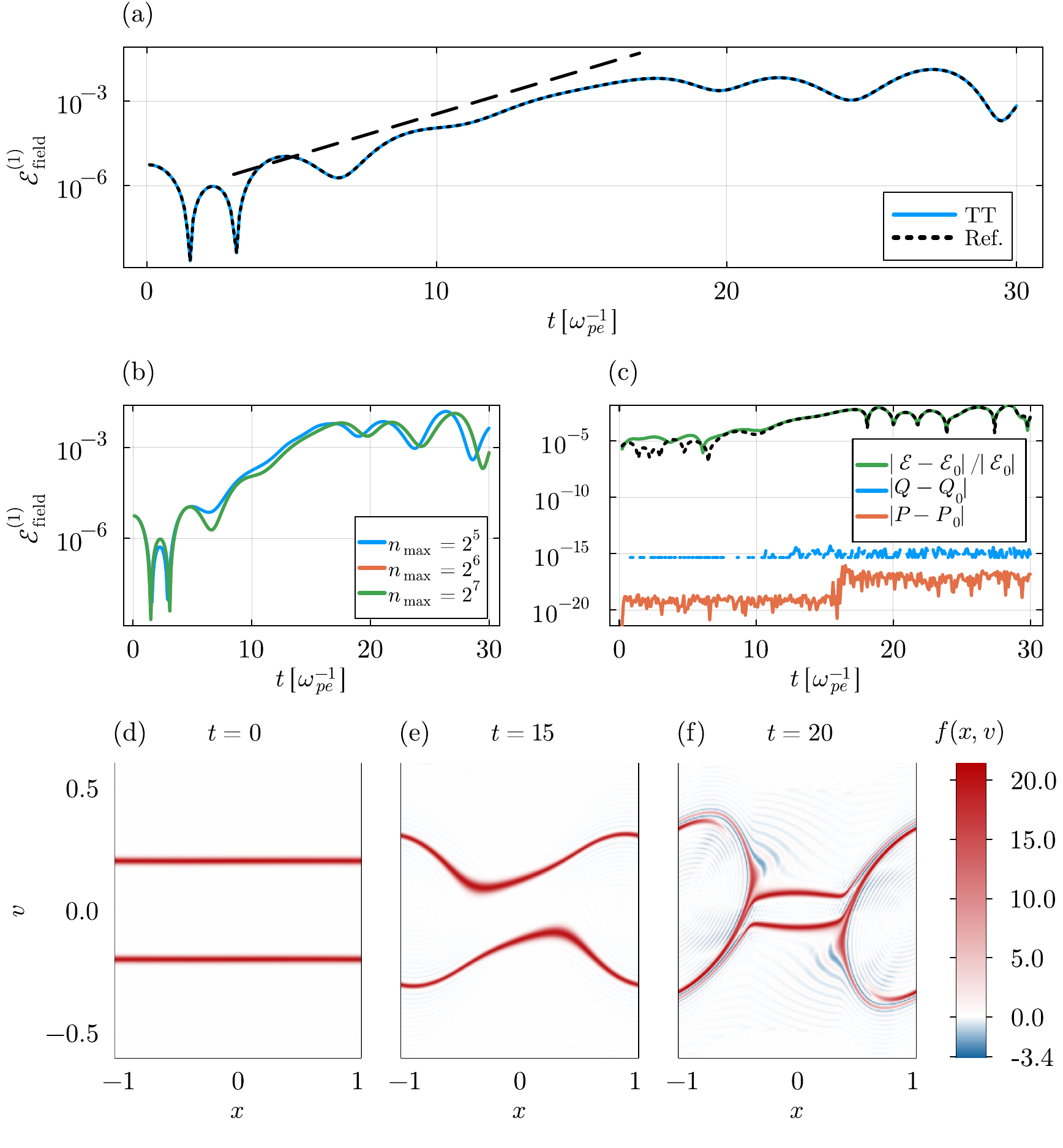}
    \caption{TT-based simulation of two-stream instability, with the initial condition set according to Eq.~\eqref{eq:ic_two_stream}.
    \textbf{(a)} Energy of the first mode of the electric field. The TT-based energy trajectory remains indistinguishable from that of the full-grid reference solver (dotted line) and accurately captures the analytic growth rate in the linear regime (dashed line).
    \textbf{(b)} Evolution of the energy of the first mode of the electric field $\mathcal{E}_\text{field}^{(1)}$ obtained for different frequency index cutoffs $n_\text{max}=2^5, 2^6, 2^7$. The $n_\text{max}=2^6$ and $n_\text{max}=2^7$ curves overlap almost exactly, so the $n_\text{max}=2^6$ curve is largely hidden in the plot.
    \textbf{(c)} Evolution of charge deviation $\lvert Q-Q_0 \rvert$, momentum deviation $\lvert P - P_0 \rvert$, and relative energy deviation $|\mathcal{E}-\mathcal{E}_0|/|\mathcal{E}_0|$ during two-stream instability. The dotted curve shows the energy-deviation trajectory of the full-grid reference. Apparent gaps in the charge-deviation curve correspond to time steps where the error is exactly zero and therefore cannot be displayed on the logarithmic y-axis.
    \textbf{(d--f)} Snapshots of the distribution function throughout the time evolution. As expected, the two streams gradually couple and eventually merge.}
    \label{fig:physical_validation_ts}
\end{figure}

The simulation results for the two-stream instability are presented in Fig.~\ref{fig:physical_validation_ts}. As shown in Fig.~\ref{fig:physical_validation_ts}(a), the growth rate of the energy of the first Fourier mode in the linear regime is in excellent agreement with the analytical prediction, confirming that the analytic growth rate of the instability is accurately captured by the solver. Compared to the full-grid reference, the TT-based solver produces a near-identical trajectory in both the linear and nonlinear regimes.

Just as for Landau damping, we study the effects of varying the frequency index cutoff $n_\text{max}$. We show the evolution of $\mathcal{E}_\text{field}^{(1)}$ obtained for $n_\text{max}=2^5, 2^6, 2^7$ in Fig.~\ref{fig:physical_validation_ts}(b). Whereas all the considered cutoff values capture the analytic growth rate within the linear regime, the smallest cutoff $n_\text{max}=2^5$ introduces a slight lag in the evolution of the field energy in the nonlinear regime, compared to the energy trajectories obtained with the higher cutoffs. The energy trajectories obtained with $n_\text{max}=2^6$ and $n_\text{max}=2^7$ remain indistinguishable throughout the full simulation, indicating that retaining modes up to index $n_\text{max}=2^6$ is sufficient for accurately resolving the dynamics with the considered simulation parameters.

The evolution of the conserved quantities is shown in Fig.~\ref{fig:physical_validation_ts}(c). Both the charge and momentum remain conserved to machine precision, reaching maximum absolute deviations from their initial values of approximately $|Q-Q_0|\sim 10^{-15}$ and $|P-P_0|\sim 10^{-17}$, respectively. Similar conservation accuracies were obtained during Landau damping, indicating that the projection-based conservation correction remains robust under different dynamics. Concerning energy conservation, the relative deviation $|\mathcal{E}- \mathcal{E}_0|/|\mathcal{E}_0|$ grows throughout the linear regime from initial errors on the order of $|\mathcal{E}- \mathcal{E}_0|/|\mathcal{E}_0|\sim 10^{-5}$ to maximum relative errors of approximately $|\mathcal{E}- \mathcal{E}_0|/|\mathcal{E}_0| \approx 10^{-2}$ upon reaching the nonlinear regime. Notably, we observe no further drift in the energy error upon entering the nonlinear regime. Although the energy conservation error of the TT-based solver differs slightly from that of the reference during the initial transient phase, both methods exhibit nearly identical error trajectories throughout the exponential growth phase and the subsequent nonlinear regime, suggesting that the TT-based representation itself imposes no additional energy conservation error.

Snapshots of the distribution function at selected times are shown in Fig.~\ref{fig:physical_validation_ts}(d)--(f), illustrating the gradual development of the instability and the eventual merging of the two streams at approximately $t = 20$. Compared to Landau damping, the negative artifacts of the distribution function are more pronounced, but remain localized, reaching a minimum value of $f_\text{min} \approx -3.4$ at $t = 20$.

\subsection{Effect of tensor-train truncation}
In this section, we study the effects of truncating the bond dimension of the TT representation of the distribution function $f$. We perform a sweep of the TT singular value cutoff $\epsilon\in \{10^{-7}, 10^{-8}, 10^{-9}, 10^{-10}\}$, keeping the remaining simulation parameters set to the values stated in Table~\ref{tab:simulation_parameters}. As before, we consider Landau damping and the two-stream instability as our test cases.

\begin{figure}
    \centering
    \includegraphics[width=.82\textwidth]{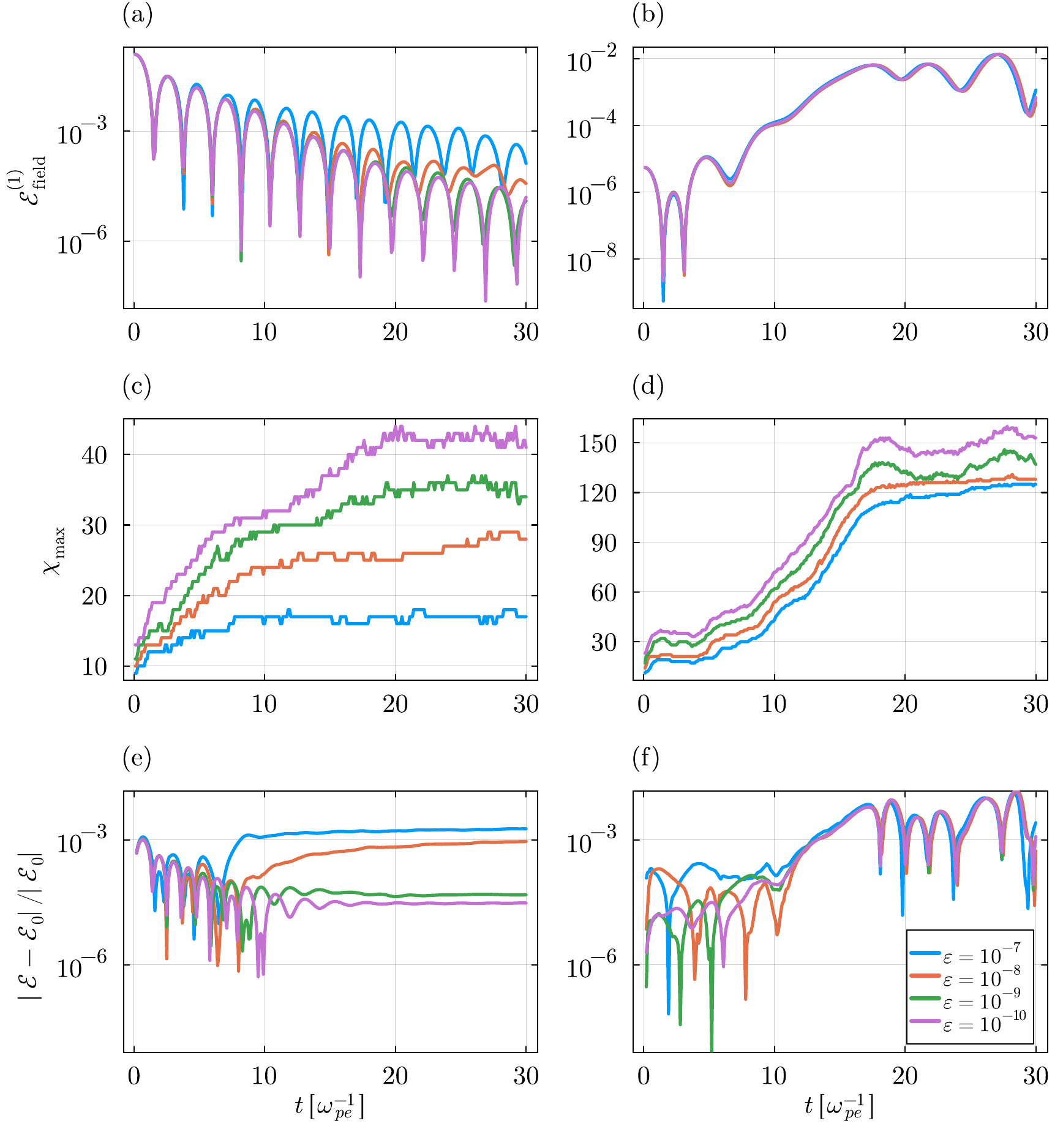}
    \caption{Effects of TT bond truncation on Landau damping and two-stream instability simulations. \textbf{(a--b)} The energy of the first mode of the electric field, $\mathcal{E}_\text{field}^{(1)}$, plotted for different values of the TT truncation cutoff $\epsilon$ for linear Landau damping and two-stream instability. For the linear Landau damping, increasing the truncation decreases the damping rate of the field energy. 
    \textbf{(c--d)} Evolution of maximum bond dimension $\chi_\text{max}$ for Landau damping and two-stream instability. For the two-stream instability, we observe a clear plateau in the bond dimension growth upon entering the nonlinear regime. \textbf{(e--f)} Conservation of energy in terms of the absolute relative energy deviation, shown separately for Landau damping and the two-stream instability.
    }
    \label{fig:truncation_effects}
\end{figure}

\begin{figure}[h]
    \centering
    \includegraphics[width=.82\textwidth]{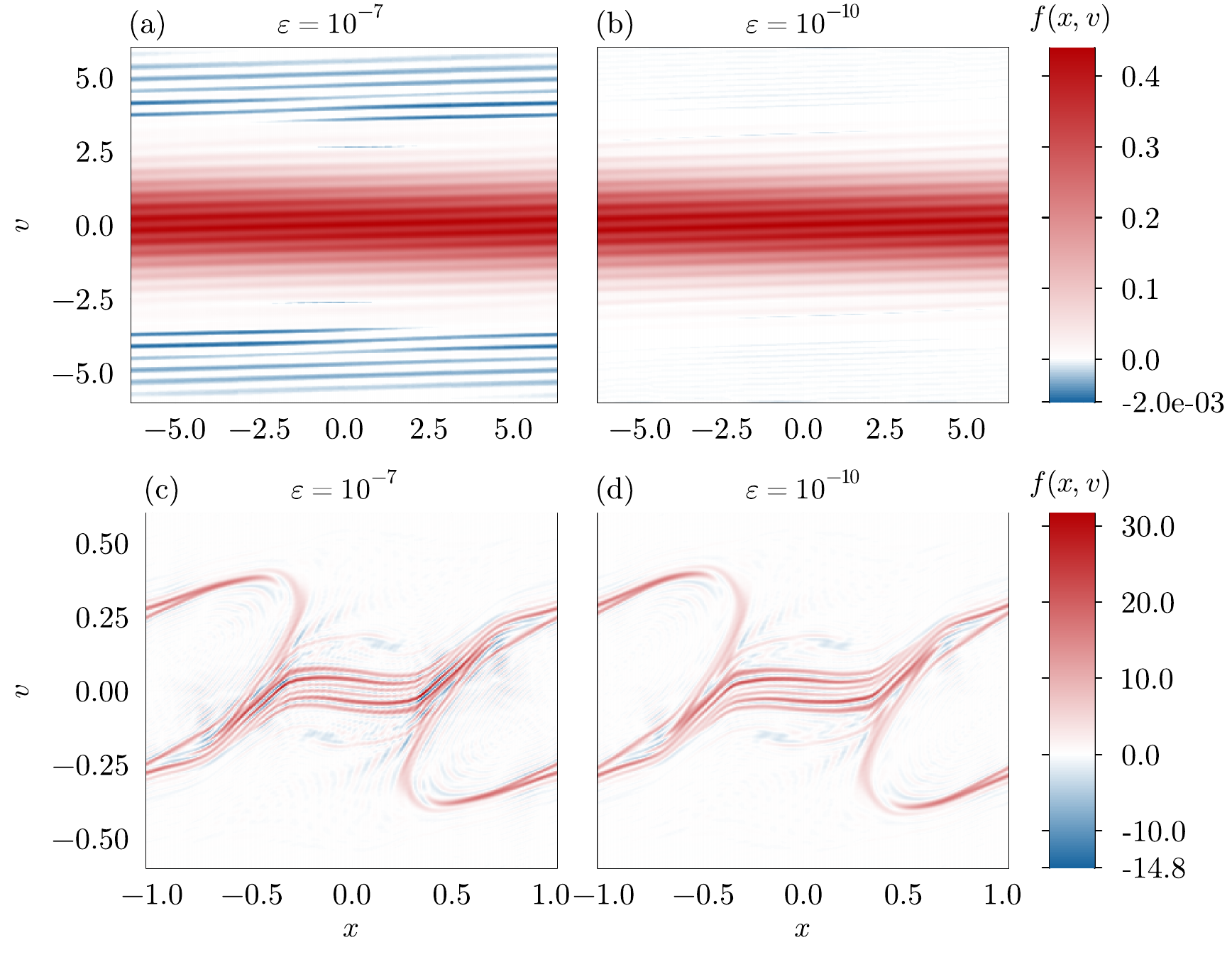}
    \caption{Distribution functions for \textbf{(a--b)} Landau damping and \textbf{(c--d)} the two-stream instability using truncation cutoff $\epsilon = 10^{-7}$ and $\epsilon = 10^{-10}$, plotted at the final simulation time $t = 30$. For both Landau damping and the two-stream instability, increasing the aggressiveness of the truncation cutoff $\epsilon$ produces larger negative artifacts.}
    \label{fig:truncation_distr}
\end{figure}

In Fig.~\ref{fig:truncation_effects}(a) and (b), we show the time evolution of the energy of the first Fourier mode of the electric field, $\mathcal{E}_\text{field}^{(1)}$, for Landau damping and the two-stream instability, respectively, as the singular value truncation threshold $\epsilon$ is varied. For Landau damping, increasing the threshold $\epsilon$ primarily manifests as a reduction in the damping rate. This effect is most pronounced for $\epsilon = 10^{-7}$, where the decay of $\mathcal{E}_\text{field}^{(1)}$ is significantly slower compared to the results obtained with more stringent truncation thresholds. In contrast, no comparable deviations in the electric field energy are observed for the two-stream instability, for which the evolution of $\mathcal{E}_\text{field}^{(1)}$ remains nearly identical across all considered cutoff values.

The maximum bond dimensions $\chi_{\text{max}}$ attained during the simulations for Landau damping and the two-stream instability are shown in Fig.~\ref{fig:truncation_effects}(c) and (d), respectively. As expected, decreasing $\epsilon$ leads to a systematic increase in $\chi_{\text{max}}$. For Landau damping, the required bond dimensions remain comparatively modest, with values in the range $15 \lesssim \chi_{\text{max}} \lesssim 45$. In contrast, the two-stream instability exhibits substantially larger bond dimensions, reaching approximately $120 \lesssim \chi_{\text{max}} \lesssim 150$ in the nonlinear regime. Notably, for the two-stream instability, the bond dimension grows steadily throughout the linear phase and then reaches a plateau once the system enters the nonlinear regime. This saturation is a favorable property, as it indicates that the computational cost associated with the TT representation does not increase indefinitely beyond the onset of nonlinear dynamics.

We study the impact of the truncation threshold $\epsilon$ on the conservation of energy in Fig.~\ref{fig:truncation_effects}(e) and (f). For Landau damping, we observe a clear reduction in the energy conservation error when decreasing the truncation cutoff $\epsilon$ from $10^{-7}$ to $10^{-9}$ after the initial fluctuations. Decreasing the cutoff further to $10^{-10}$ results in only a minor improvement in energy conservation, suggesting that the remaining conservation error originates from sources other than bond truncation. For the two-stream instability, the energy conservation is largely unaffected by the choice of $\epsilon$. This is particularly the case within the nonlinear regime, where we obtain near-identical energy errors, regardless of the choice of $\epsilon$.

Concerning the phase-space distributions, the observed negative numerical artifacts exhibit a clear and systematic dependence on the truncation level $\epsilon$, indicating that they are directly linked to the low-rank compression of the TT representation. In particular, more aggressive truncation makes the negativity more pronounced, as can be observed in Fig.~\ref{fig:truncation_distr}. For Landau damping, increasing the truncation threshold from $\epsilon=10^{-10}$ to $\epsilon=10^{-7}$ increases the magnitude of the negative undershoots by roughly one order of magnitude, with the minimum value changing from $f_{\min}\approx -3.6\times10^{-4}$ to $f_{\min}\approx -2\times10^{-3}$. A similar trend is observed for the two-stream instability, where the minimum value shifts from $f_{\min} \approx -5$ to $f_{\min} \approx -15$ when varying $\epsilon$ in the same range. Importantly, these large negative values remain strongly localized to small, isolated regions of phase space rather than being distributed globally. Strategies for mitigating negative artifacts and ensuring positivity preservation within the TT framework are discussed in Section~\ref{sec:discussion}.

\subsection{Runtime and compression analysis}
\begin{figure}[ht]
    \centering
    \includegraphics[width=.82\textwidth]{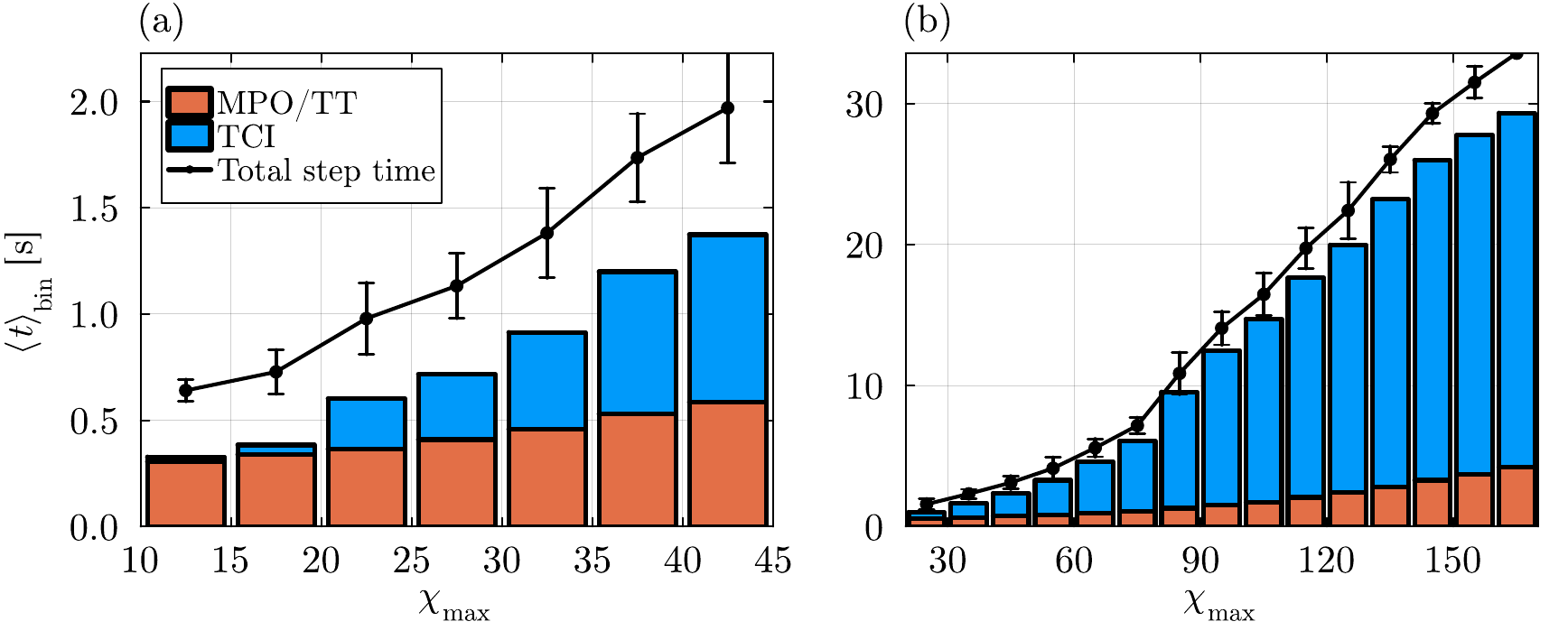}
    \caption{
    Average per-step runtime for the TCI acceleration step and TT--MPO contractions in simulations of \textbf{(a)} Landau damping and \textbf{(b)} the two-stream instability, using a bond-truncation cutoff of $\epsilon = 10^{-10}$. The black line indicates the mean total per-step runtime ($\pm$ standard deviation).
    }
    \label{fig:runtime-histogram}
\end{figure}

Finally, we perform an analysis of the algorithm runtime and compression ratios obtained via the TT representation. In Fig.~\ref{fig:runtime-histogram} we show the average runtime per time step for the Landau and two-stream simulations. The runtime is decomposed into the contribution associated with the TCI acceleration step and the MPO-TT contractions (including the subsequent bond truncations). Although the MPO-TT contraction steps dominate the runtime at small bond dimensions of $\chi_\text{max} \lesssim 30$, TCI quickly becomes the dominant runtime contribution at larger bond dimensions. This is most pronounced in the two-stream case, where the substantially larger bond dimensions required for accurate state representation lead to a significant increase in the TCI runtime. 

Although larger bond dimensions generally lead to longer runtimes, the measured wall times do not appear to follow a simple power-law dependence on the observed maximum bond dimension $\chi_\mathrm{max}$. This is not surprising, since $\chi_\mathrm{max}$ provides only a coarse summary of the TT state and does not capture the full bond-dimension distribution across the TT chain. Furthermore, as the TCI acceleration step is warm-started from the TT representation obtained at the previous time step, the number of sweeps and TCI queries required for convergence depends on the similarity between successive states. Consequently, time steps with comparable values of $\chi_\mathrm{max}$ may still exhibit noticeably different runtimes.

\begin{figure}[h]
    \centering
    \includegraphics[width=.82\textwidth]{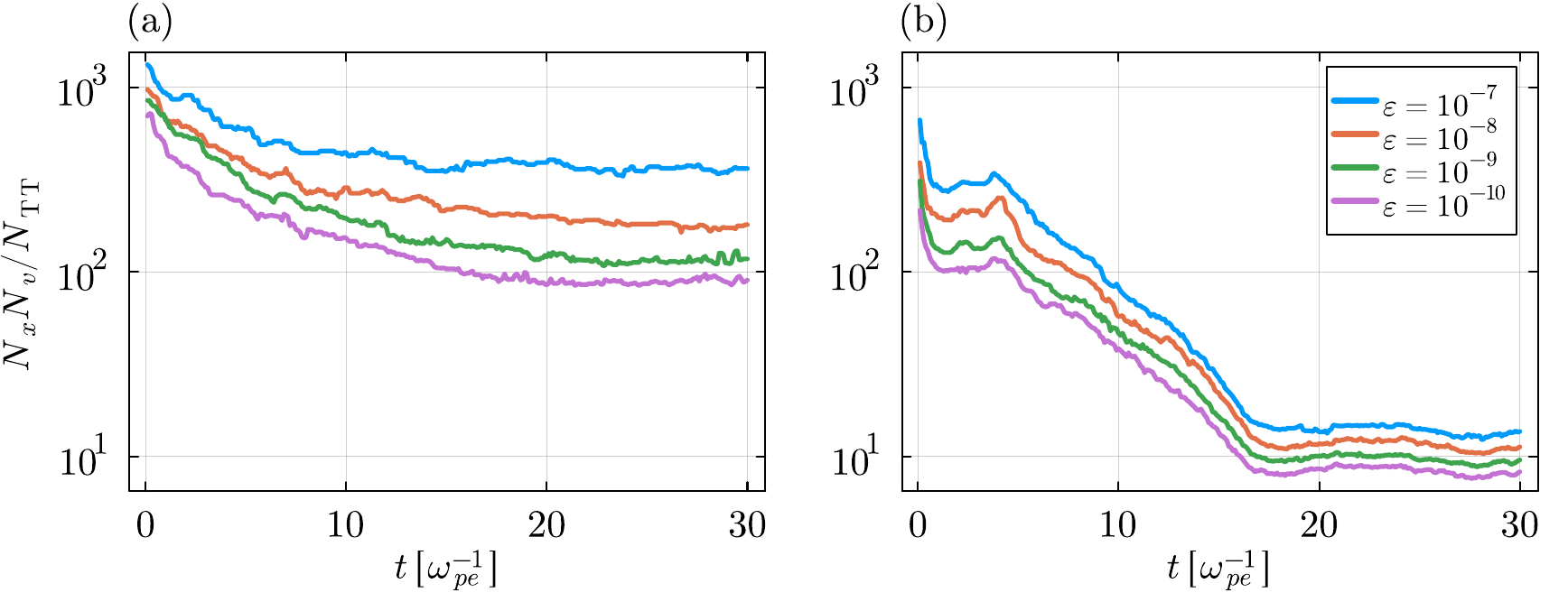}
    \caption{
    The compression ratio $N_x N_v/N_\text{TT}$ between the full-grid and compressed TT-based representations. The compression ratio is displayed separately for \textbf{(a)} Landau damping and \textbf{(b)} two-stream instability for varying values of the TT bond truncation threshold $\epsilon$.
    }
    \label{fig:compression-ratio}
\end{figure}

While the runtime analysis highlights the computational overhead associated with TCI reconstruction and TT arithmetic, the primary advantage of the TT representation lies in its ability to compress the phase-space distribution. To quantify the degree of compression obtained throughout the simulations, we consider the \textit{compression ratio}, defined as the ratio between the number of points $N_x N_v$ in the full-grid representation and the number of entries $N_\text{TT}$ in the TT-state representation. Fig.~\ref{fig:compression-ratio} shows the compression ratio $N_x N_v/N_\text{TT}$ throughout the Landau damping and two-stream reference simulations. For Landau damping, the final-time compression ratio ranges from $90$ to $360$, depending on the truncation threshold $\epsilon$. For the two-stream simulations, the degree of compression varies substantially between the linear and nonlinear regimes. Whereas the initial phase exhibits compression ratios exceeding two orders of magnitude, the final nonlinear-regime values range from $8$ to $14$, depending on $\epsilon$. Although the achievable compression depends strongly on the complexity of the underlying phase-space dynamics, these results demonstrate that the TT representation can reduce the memory footprint by up to several orders of magnitude relative to the full-grid representation while maintaining excellent agreement with the reference solution.

\section{Discussion}
\label{sec:discussion}
The results presented in this work indicate that a combination of a quantics TT-based representation of the distribution function and a spectral solver built around the low-rank MPO representation of the Fourier transform yields accurate results and successfully reproduces several important features characteristic of collisionless plasmas. In particular, the method reproduces the expected damping and growth rates of the benchmark problems while achieving substantial compression relative to the corresponding full-grid representation. Beyond the numerical performance demonstrated here, the proposed framework illustrates how Fourier transforms with highly compressible tensor representations can be incorporated directly into tensor-network plasma solvers as computational primitives. This perspective suggests several avenues for future work, including adaptive spectral filtering and hybrid physical-space/Fourier-space tensor-network schemes.

A limitation of the proposed method is the emergence of negative numerical artifacts in the distribution function, which become increasingly pronounced under more aggressive bond dimension truncation. This is a common issue in Vlasov solvers, which is usually addressed via positivity-preserving schemes~\cite{kiechle_positivity-preserving_2025}. One promising direction for mitigating this issue, within the TT framework, is to represent the square root of the distribution function, $\sqrt{f}$, rather than $f$ itself. This idea has recently been proposed in a non-TT context as a means of enforcing positivity by construction~\cite{issan_anti-symmetric_2024}. From a conceptual standpoint, such a formulation is closer in spirit to quantum simulations and quantum-inspired tensor network methods, where one represents an underlying amplitude while physically relevant observables are obtained only after taking suitable quadratic forms. Moreover, since SVD-based truncation is optimal with respect to the $L^2$ norm, whereas the physically relevant norm for $f$ is the $L^1$ norm, a square-root formulation may lead to truncations that more faithfully preserve key physical properties of the system.

Concerning performance, we found that the TCI step used to fit the accelerated distribution dominated the overall runtime. This is partially explained by the intrinsic complexity of the interpolation task, which requires fitting a TT representation over the combined position and velocity registers. In addition, each TCI query of the accelerated state is computationally expensive, as it involves evaluating both the electric field $E(x_{i_x})$ and the distribution function $f(x_{i_x}, k_{i_v})$ through separate TT contractions. It is likely that this performance bottleneck could be partially mitigated through shared-memory parallelization strategies~\cite{dolgov_parallel_2020}, parallel evaluation of TCI queries, and a systematic investigation of TCI algorithmic variants, including single- and two-site updates and alternative pivot-exploration strategies~\cite{fernandez_learning_2025}. Furthermore, it may be worthwhile to consider alternative bit-ordering schemes beyond the interleaved ordering, which might better accommodate low-rank structure in the phase-space distribution. This question is particularly relevant in the present setting, as the MPO representation of the Fourier transform reverses the quantics bit order of the transformed register, thereby breaking the original ordering of the position and velocity registers.

A natural extension of the present framework is the treatment of higher-dimensional systems. At the representation level, a $D$-dimensional distribution function $f(\mathbf{x},\mathbf{v})$ can be incorporated into the TT format by introducing additional quantics bit-registers for each spatial and velocity component, increasing the TT order from $2R$ to $2DR$. In this setting, the ordering of the registers becomes an additional design choice, since different interleaving strategies may affect how efficiently correlations between coordinates and scales are captured. The main algorithmic components of the solver also generalize in a fairly direct way: Fourier transforms can still be applied register-wise through TT--MPO contractions, the advection step remains separable across coordinate pairs $(x_j,v_j)$, and charge density is still obtained by contracting over the velocity registers. Solving for the electric field becomes more involved, since after forming $\rho(\mathbf{x})$, one must compute each field component $E_j(\mathbf{x})$ by applying a multidimensional Fourier-space multiplier proportional to $- i k_j/|\mathbf{k}|^2$, rather than the simple one-dimensional kernel $-i/k$. Likewise, the acceleration step would require TCI reconstruction in a larger tensor domain, with each query involving evaluations of both the TT state and multiple electric-field components. Thus, the main open question in higher dimensions is not whether the split-operator steps can still be formulated in TT/MPO form, but whether the required tensor objects remain sufficiently compressible for the overall method to remain efficient.

Before concluding, it is useful to place the TT truncation employed here in the broader context of closures used throughout plasma physics to render complex kinetic descriptions tractable. In kinetic theory, the BBGKY hierarchy organizes particle correlations, and practical models such as the Vlasov or Boltzmann equations are obtained by closing the hierarchy under assumptions appropriate to a given regime~\cite{ichimaru_statistical_plasma_2004,balescu_charged_1963}. Similarly, fluid and MHD models follow from taking velocity moments of the kinetic equations and introducing closures that truncate the resulting infinite moment hierarchy~\cite{grad_kinetic_1949,levermore_moment_1996}. The quantics TT formulation introduces a different notion of closure. Rather than truncating particle correlations or moment equations, TT truncation limits complexity by restricting the rank of a tensorized representation of $f$, thereby weakening inter-scale couplings in the quantics hierarchy. Practically, this is realized by discarding low-weight singular values, which adaptively compresses the state while preserving its dominant separable structure. In this sense, TT truncation serves an analogous purpose to classical closures, controlling complexity while acting on the numerical representation instead of imposing an explicit physical approximation.

\section{Conclusion}
\label{sec:conclusion}
In this work, we have presented a spectral solver for the collisionless Vlasov--Poisson system in which the phase-space distribution function $f$ is represented in quantics tensor train format and advanced directly in compressed form. A central feature of the method is that spectral operations, including the Fourier transforms required for time stepping, are carried out directly within the compressed representation via efficient tensor contractions. The self-consistent electric field is likewise obtained entirely within the same framework, through tensor contractions that extract the required velocity moments and a spectral Poisson solver, without reconstructing $f$ on a full grid. In this way, the method addresses a main bottleneck of Eulerian kinetic solvers, namely the storage and manipulation of high-resolution phase-space data. We validated the method on standard benchmarks, including linear Landau damping and the two-stream instability, where the TT solver remained in near-indistinguishable agreement with the full-grid reference and reproduced the expected damping and growth behavior of the electric field. Beyond benchmark verification, we systematically investigated how varying the TT truncation threshold, and thereby the degree of compression, affects conservation properties, positivity behavior, and overall computational cost. Notably, the TT representation yielded final-time compression ratios ranging from approximately 90 to 360 for Landau damping and 8 to 14 for the two-stream instability, with substantially larger compression attained during the initial linear phase.

At the same time, important open questions remain. In particular, further work is needed to characterize truncation effects more comprehensively across longer integration times and more strongly nonlinear regimes, and to develop diagnostics that reliably predict when compressibility deteriorates~\cite{kusch_stability_2023,einkemmer_review_2025}. In addition, in our current implementation, the repeated TCI reconstructions required for the acceleration step constitute the dominant computational cost, motivating future work on more efficient update strategies, improved interpolation and query policies, and performance optimizations~\cite{savostyanov_oseledets_fast_2011, kusch_second-order_2025}. Extending the framework to higher-dimensional settings and to the Vlasov--Maxwell system, while retaining an efficient low-rank representation of the evolving state, is another natural direction for future work~\cite{einkemmer_low-rank_2020}. Despite these open challenges, our results support the view that tensor network compression can be used not only for storage but also as a computational representation in which kinetic equations are advanced directly, making spectral tensor network solvers a promising route toward high-accuracy Eulerian Vlasov simulations.

\section{Data availability}
The code and data used to generate the numerical results presented in this work are available in the public repository \texttt{VlasovTT.jl}~\cite{asgrim2026_VlasovTT}.

\section*{Acknowledgments}
This work is funded by the European Union. This work has received funding from the European High Performance Computing Joint Undertaking (JU) and Sweden, Finland, Germany, Greece, France, Slovenia, Spain, and the Czech Republic under grant agreement No. 101093261, Plasma-PEPSC.

\bibliographystyle{elsarticle-num}
\bibliography{cas-refs}

\newpage
\appendix

\end{document}